\documentclass[twocolumn,aps,prl,groupedaddress,amsmath,amssymb]{revtex4-1}
\usepackage{url}
\usepackage[colorlinks=true, linkcolor=blue,urlcolor=blue,anchorcolor=blue,citecolor=blue,bookmarksnumbered]{hyperref}
\usepackage{graphicx}
\usepackage{dcolumn}
\usepackage{bm}
\usepackage{verbatim}
\usepackage{ulem}

\begin{document}
	
	\title{Cascaded Rydberg antiblockade: Multi-atom excitation dynamics and entanglement}
	\author{Jin-Lei~Wu$^{1}$}
	\author{Jun~Wu$^{2}$}
	\author{Pei-Yao~Song$^{1}$}
	\author{Yan~Wang$^{3}$}
	\author{Ya~Gao$^{1*}$}
	\author{Xue-Ke~Song$^{2*}$}
	\author{Shi-Lei~Su$^{1,4*}$}
	\affiliation{$^{1}$School of Physics, Zhengzhou University, Zhengzhou 450001, China\\
		$^{2}$School of Physics, Anhui University, Hefei 230601, China\\
		$^{3}$School of Electronics and Information, Zhengzhou University of Light Industry, Zhengzhou 450001, China\\
		$^{4}$Institute of Quantum Materials and Physics, Henan Academy of Science, Henan 450046, China}
	
		\affiliation{\medskip*Corresponding authors (Ya~Gao, email: ygao@zzu.edu.cn; Xue-Ke~Song, email: songxk@ahu.edu.cn; Shi-Lei~Su, email: slsu@zzu.edu.cn)}

	\begin{abstract}
		\noindent We propose a cascaded Rydberg antiblockade~(RAB) regime via a Floquet modulation in four fully connected interacting atoms, which establishes a new synthetic dimension, Dicke-state lattice~(DSL), in the space of collective spin excitations. By applying a global periodic driving, we synthesize an effective Hamiltonian that enables perfect state transfer across the five-site DSL with multiple programmable pathways from stepwise nearest-neighbor jumps to a single-step transition. This DSL platform further allows us to simulate a dynamic Su-Schrieffer-Heeger model, where soft quantum control is employed to achieve topologically inspired full RAB $|0000\rangle \to |1111\rangle$ with enhanced robustness against disorder. Moreover, by incorporating the shortcut to adiabaticity technique, we generate high-fidelity entangled twin-Fock and Greenberger-Horne-Zeilinger states on the four atoms within sub-microsecond timescales, outperforming the speed limits of conventional adiabatic protocols. Our work demonstrates a flexible and programmable synthetic dimension for quantum simulation and multipartite entanglement engineering in Rydberg atom arrays, paving the way for the future development of quantum information processing.\\
		\\
		\noindent{\bf Rydberg atoms, Floquet modulation, synthetic dimension, shortcuts to adiabaticity}
	\end{abstract}
	\maketitle
	\noindent \textbf{\begin{large}1~~~{Introduction}\end{large}}\\
	\\
	The precise control of collective spin excitation dynamics in interacting many-body systems lies at the heart of modern quantum science~\cite{JZhang2017Nature,JYao2021PRX,YYao2023NP,QWang2025SCPMA,YTLee2026PRL}. Among the various quantum platforms, neutral atoms excited to Rydberg states have emerged as one of the most promising candidates for studying and controlling interacting spin dynamics~\cite{Saffman2010RMP,Browaeys2020NP,ZYZhang2020,Morgado2020AVSQS,XLWu2021CPB,XFShi2022QST,XQShao2024APR,SXLi2024PRA,HYTang2024APLQuantum}. The celebrated Rydberg blockade mechanism due to strong Rydberg-Rydberg interactions (RRI), where the excitation of a single atom to a Rydberg state inhibits further excitations in its vicinity, has been widely exploited for implementing high-fidelity quantum gates and generating entanglement \cite{YZeng2017PRL,ZYAn2025PRL,FromonteilPRXQuantum2023,Evered2023Nature,SLSu2024CPL,YSun2024SCPMA,CFSun2024EPJQT}. Conversely, the Rydberg antiblockade (RAB) regime, in which multiple atoms can be simultaneously excited into interacting Rydberg states despite strong interactions, provides a natural and efficient pathway for parallel preparation of multipartite entangled states~\cite{Ates2007PRL,thm2012,Gambetta2020PRL,SLSu2020EPL,FLiu2022PRL,XXLi2024PRA,FQGuo2025PRA}. However, achieving coherent and programmable control over collective excitation manifolds of multiple atoms remains challenging, as it requires precise engineering of the spectral structure and transition pathways between collective spin states.
	
	Periodic driving or Floquet engineering has proven to be a powerful tool for tailoring effective Hamiltonians in quantum systems~\cite{Eckardt2017RMP,Nguyen2024NP,Zhou2024NC}. In the context of Rydberg atoms, Floquet methods have been used to enhance blockade efficiency~\cite{Basak2018PRL,Zhao2023NC,Mallavarapu2021PRA,YWei2026PRA}, realize two-photon RAB transitions~\cite{JLWu2020OL,JLWu2021PRJ}, construct robust quantum gates~\cite{JLWu2021PRA,JLWu2022QUTE,JWu2025npjQI}, and accelerate entanglement generation~\cite{RLi2020PRA,JLWu2021PRApplied}. When combined with the RAB condition, periodic modulations offer a route to engineer cascaded transitions between Dicke states, constructing a one-dimensional lattice in the space of excitation numbers, defining a new synthetic dimension termed the Dicke-state lattice~(DSL). This DSL serves as a programmable playground for studying controlled spin excitation dynamics, analogous to a tight-binding chain where each site represents a distinct collective spin configuration.
	
	\begin{figure*}[t]
		\centering
		\includegraphics[width=0.96\linewidth]{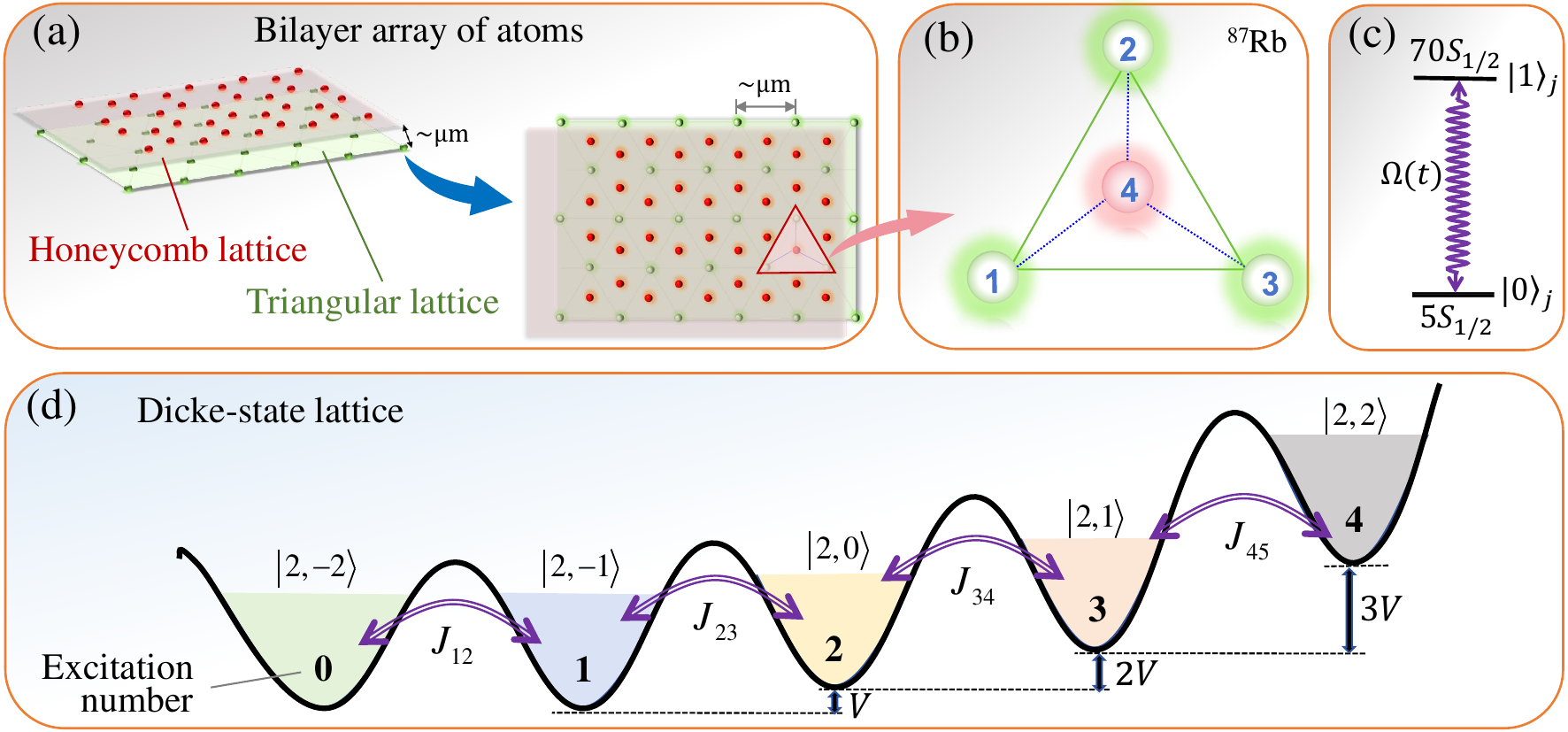}
		\caption{\label{fig1}(a)~Schematic diagram of a bilayer neutral atom array, with one layer holding single atoms assembled at sites in a triangular optical lattice while the other in a honeycomb lattice. (b)~Four fully connected atoms with an equidistant interaction strength $V$ between arbitrary two. (c)~Atomic spin transition being identically driven by a periodic amplitude-modulated field, from the ground state $|0\rangle$ to the Rydberg state $|1\rangle$ with a Floquet Rabi frequency $\Omega(t)$. (d)~One-dimensional synthetic dimension in DSL with the individually addressable nearest-neighbor hopping rates.}
	\end{figure*}
	In this work, we propose and systematically investigate a cascaded RAB regime in four fully connected interacting atoms via Floquet modulations. This regime enables first-order dynamics significantly faster than conventional higher-order RAB processes~\cite{SLSu2018PRA,THXing2020PRA} and requires only global single-shot operations, offering a more straightforward route than stepwise or addressing-based schemes~\cite{SLSu2017PRA_fast,SYBai2020NJP}.	
	By applying a global driving field, we derive an effective Hamiltonian that synthesizes a hopping-addressable five-site DSL with tunable hopping rates, building a new synthetic dimension~\cite{Boada2012PRL,Ozawa2019NRP,Ehrhardt2023LPR,Argueello2024CP,DYu2025PI}. We demonstrate perfect state transfer~(PST) across the DSL, revealing multiple programmable pathways ranging from nearest-neighbor hopping to direct long-range transfers. Furthermore, we simulate a dynamic Su-Schrieffer-Heeger~(SSH) model~\cite{SSH1979PRL} and employ soft quantum control~\cite{Haase2018PRL} to achieve topologically inspired full RAB ($|0000\rangle \to |1111\rangle$) with enhanced robustness against disorder. Finally, using shortcut-to-adiabaticity~(STA) methods~\cite{STA2019RMP}, we generate high-fidelity four-particle twin-Fock and Greenberger-Horne-Zeilinger~(GHZ) states within sub-microsecond timescales. Our results establish a versatile synthetic dimension for quantum simulation and multipartite entanglement engineering in Rydberg atoms, providing an expandable research direction for multiparticle quantum information processing.\\
	\\

	\noindent \textbf{\begin{large}2~~~{Hamiltonian and Dicke-state lattice}\end{large}}\\
	\\
	As shown in Figure~\ref{fig1}(a), we consider a neutral atom system involving four fully connected interacting Rydberg atoms on vertices of a tetrahedral structure [Figure~\ref{fig1}(b)] trapped in a bilayer atomic array~\cite{Ywang2015PRL,YWang2016Science,Barredo2018Nature,Kumar2018Nature,ZMMeng2023Nature}, one layer in a triangular optical lattice while the other in a honeycomb lattice. This specific three-dimensional architecture is strictly required by the cascaded RAB condition. To ensure that the interaction strengths between any two atoms are identical ($V_{jk} = V$), the four atoms must be mutually equidistant. Since a 2D plane can accommodate at most three mutually equidistant points (an equilateral triangle), a single-layer configuration (such as a 2D Kagome lattice) is geometrically insufficient. By utilizing such a bilayer configuration with an appropriately tuned interlayer distance, regular tetrahedral clusters can be naturally constructed. The four identical atoms are pumped by a laser inducing a transition between a ground state $|0\rangle$ and a Rydberg state $|1\rangle$ with a modulated Rabi frequency $\Omega(t)$, as shown in Figure~\ref{fig1}(c). From an experimental perspective, the realization of such a highly symmetric tetrahedral geometry is feasible. Deterministic 3D atomic assembly has been demonstrated using holographic tweezers \cite{Barredo2018Nature} and optical lattices \cite{Kumar2018Nature}. To ensure uniform interaction $V_{jk}$, we employ the $70S_{1/2}$ state, which exhibits an intrinsically isotropic van der Waals~(vdW) interaction ($V \sim1/d^6$) due to its spherical symmetry, making it independent of geometric orientation \cite{Saffman2010RMP,XQShao2024APR}. Furthermore, by applying a global uniform magnetic field to define a common quantization axis and utilizing a laser beam waist significantly larger than the tetrahedral extent, all atoms experience identical laser intensity and polarization, resulting in highly uniform coherent driving \cite{Levin2018PRL,Evered2023Nature}.
	
 The Hamiltonian can be written in the interaction picture ($\hbar\equiv1$)
	\begin{equation}\label{eq1}
	\hat{H}(t)=\sum_{j=1}^4\frac{\Omega(t)}{2}\hat{\sigma}_x^j + \sum_{j>k=1}^4 V_{jk}|11\rangle_{jk}\langle 11|,
	\end{equation}
	where $\hat{\sigma}_x^j=\sigma_-^j+\sigma_+^j$, with $\sigma_-^j=(\sigma_+^j)^\dagger=|0\rangle_{j}\langle 1|$, is the Pauli-$x$ operator, and $V_{jk}$ denotes the vdW RRI strength between the $j$th and $k$th atoms. Here, we adopt Floquet-modulated pulse engineering and set $\Omega(t)$ as an even function of a truncated Fourier series~\cite{Claeys2019PRL,JLWu2022QUTE,HZhou2019SB}
	\begin{equation}\label{eq2}
	\Omega(t)=\sum_{n=0}^N \Omega_n(t) \mathrm{cos}(n\omega t),
	\end{equation}
	where $\Omega_n(t)$ and $n\omega$ are the amplitude component and the modulation frequency, respectively, of the $n$th-order element. The undetermined value of $N$ can be chosen as needed~($N=3$ in this work). Physically, this effective time-dependent Rabi frequency $\Omega(t)$ is realized via a two-photon excitation scheme ($|0\rangle \rightarrow |p\rangle \rightarrow |1\rangle$). To minimize spontaneous emission from the intermediate state $|p\rangle$, the multi-frequency amplitude modulation is applied exclusively to the upper transition ($|p\rangle \leftrightarrow |1\rangle$) using an arbitrary waveform generator (AWG) and an acousto-optic modulator (AOM), while the lower transition ($|0\rangle \leftrightarrow |p\rangle$) is kept constant. A rigorous derivation of the effective Hamiltonian via adiabatic elimination, along with the compensation mechanism for the accompanying time-dependent AC Stark shifts, is detailed in Appendix B.
	
	We set the RAB condition $V_{jk}= V=\omega$ in the Rydberg blockade regime $V\gg\max[\Omega_n(t)]$ and turn to the rotation frame with respect to $\sum_{j>k=1}^4 V_{jk}|11\rangle_{jk}\langle 11|$~\cite{JLWu2022QUTE}. After neglecting the off-resonant terms with large detunings, an effective cascaded RAB Hamiltonian, describing the nearest-neighbor hopping in a five-site DSL shown in Figure~\ref{fig1}(d), is obtained (see Appendix~A for detailed derivation)
	\begin{equation}\label{eq3}
	\hat{H}_{\mathrm{eff}}(t)=\sum_{n=0}^3J_{n+1,n+2}(t)|2,n-2\rangle \langle 2,n-1|+\mathrm{H.c.},
	\end{equation}
	with the DSL hopping rate $J_{n+1,n+2}(t)\equiv\lambda_n\Omega_n(t)/[4-2\delta(n)]$.
	$\lambda_n\equiv\sqrt{(n+1)(4-n)}$ is the collective enhancement factor of the four-spin transitions. $\delta(n)$ is the Dirac delta function. The five fully symmetric Dicke states of four spins are defined as
	\begin{eqnarray}\label{eq4}
	|2,-2\rangle&=&|0000\rangle,\nonumber\\
	|2,-1\rangle&=&\frac{1}{2}(|1000\rangle+|0100\rangle+|0010\rangle+|0001\rangle),\nonumber\\
	|2,0\rangle&=&\frac{1}{\sqrt{6}}(|1100\rangle+|1010\rangle+|1001\rangle+|0110\rangle+|0101\rangle\nonumber\\
	&&+|0011\rangle),\nonumber\\
	|2,1\rangle&=&\frac{1}{2}(|1110\rangle+|1101\rangle+|1011\rangle+|0111\rangle),\nonumber\\
	|2,2\rangle&=&|1111\rangle.
	\end{eqnarray}
	These states are eigenvectors of the four-atom RRI Hamiltonian with corresponding eigenenergies
	\begin{equation}\label{eq5}
	E_0=0,\quad E_1=0,\quad E_2=V,\quad E_3=3V,\quad E_4=6V.
	\end{equation}\\
	\\

	\begin{figure}[t]
		\centering
		\includegraphics[width=\linewidth]{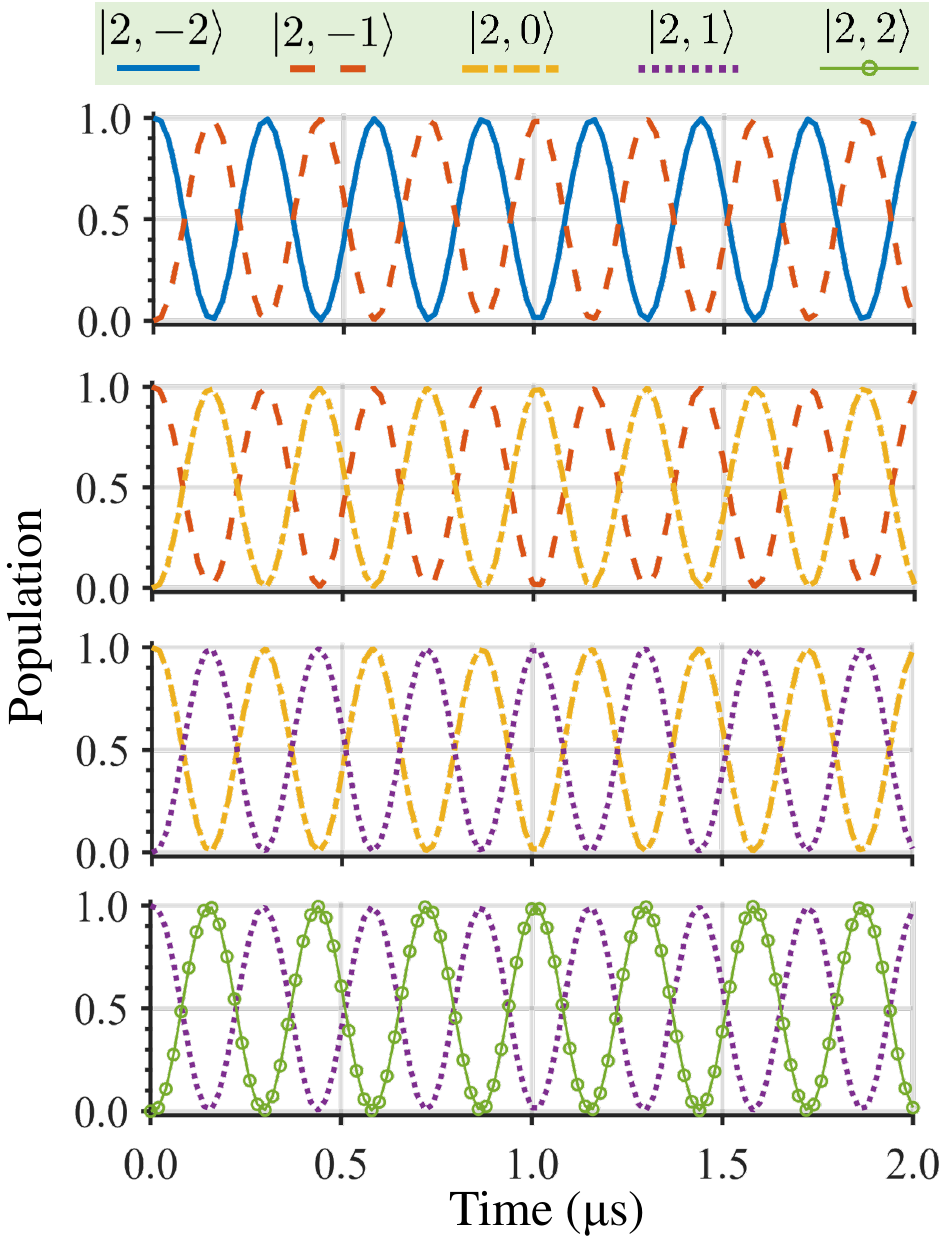}
		\caption{\label{fig2}~Complete Rabi-like oscillations between two adjacent DSL sites $|2,n-2\rangle$ and $|2,n-1\rangle$ with the initial state $|2,n-2\rangle$, with $n$ increasing from the uppermost panel $n=0$ to the lowest $n=3$, respectively.}
	\end{figure}
	\noindent \textbf{\begin{large}3~~~{Rabi oscillations and state transfer}\end{large}}\\
	\\
To validate our DSL Hamiltonian in eq.~\eqref{eq3}, we first investigate Rabi-like oscillations between each pair of the nearest-neighbor sites, for which the $l$-th site population $P_{l}(t)={\rm trace}[\hat\rho(t)|2,l-3\rangle\langle 2,l-3|]$ is obtained through numerically solving master equation
\begin{equation}\label{eq8}
\frac{\partial\hat\rho(t)}{\partial t}=-i[\hat{H}(t),\hat\rho(t)]+\sum_{j=1}^4{\mathcal{\hat L}}_j[\hat\rho(t)],
\end{equation}
with $\hat\rho(t)$ representing the density operator and $\hat{H}(t)$ the full Hamiltonian in eq.~\eqref{eq1}. The Lindblad operator ${\mathcal{\hat L}}_j[\hat\rho(t)]\equiv0.5\Gamma[2\sigma_-^j\hat\rho(t)\sigma_+^j-\sigma_+^j\sigma_-^j\hat\rho(t)-\hat\rho(t)\sigma_+^j\sigma_-^j]$ describes the spontaneous decay effect, where $\Gamma \equiv 1/\tau$ is the atomic decay rate with $\tau$ being the lifetime of the Rydberg state. The laser pump process from the ground state $|0\rangle$ to the Rydberg state $|1\rangle$ can be implemented via a two-photon transition in $^{87}\text{Rb}$ atoms (see Appendix~B for more details)~\cite{Evered2023Nature,Levin2018PRL,Levin2019PRL}. Accordingly, we consider the hyperfine ground states $|0\rangle = |5S_{1/2}, F = 1, m_F = 0\rangle$, an intermediate state $|p\rangle = |6P_{3/2}\rangle$, and the Rydberg state $|1\rangle = |70S_{1/2}\rangle$, which features an interaction coefficient of $C_6/2\pi = 858.4$~GHz~$\mu\text{m}^6$. At an experimentally accessible atomic temperature of $T_a = 10\ \mu\text{K}$, the lifetime $\tau$ of the $|70S_{1/2}\rangle$ state in $^{87}\text{Rb}$ is approximately $400\ \mu\text{s}$~\cite{Evered2023Nature,Levin2018PRL,Levin2019PRL}. Here we set an interatomic separation of $d = 4.8~\mu\text{m}$, the corresponding RRI strength is $V = 2\pi \times 70.18\ \text{MHz}$. The hopping rate $J_{n+1,n+2}(t)$ is considered time-independent as $J_{n+1,n+2}=\Omega=2\pi\times1.75$~MHz between the DSL sites $|2,n-2\rangle$ and $|2,n-1\rangle$, and the amplitude components in $\Omega(t)$ are obtained by $\Omega_{n'=n}=\Omega[4-2\delta(n)]/\lambda_{n}$ and $\Omega_{n'\neq n}=0$. In Figure~\ref{fig2}, the complete Rabi-like oscillations between arbitrary two adjacent DSL sites are observed, indicating validity of our approximation for DSL depicted in eq.~\eqref{eq3}.

Furthermore, closely related to the continuous-time quantum walk~\cite{Farhi1998PRA}, it is one of the central tasks to evolve the system to a certain Dicke state in quantum information processing (to push the particle into a specified site in quantum simulation), especially as a matter of fact that $|2,-1\rangle$, $|2,0\rangle$, $|2,1\rangle$ are fully entangled states among the four atoms. In Figure~\ref{fig3}, we show four ways, attributed to
tunability and flexibility of our DSL, to push
a pseudo-particle from the first site $|2,-2\rangle$ to the fifth $|2,2\rangle$ by using the condition of PST~\cite{Christandl2004PRL,XLi2018PRApplied}. In Figure~\ref{fig3}(a), the particle hops to the fifth site by four steps with three times of quenched Hamiltonian, where the parameters are set as
\begin{eqnarray}
&& \Omega_{0}=\Omega,\quad \Omega_{1,2,3}=0,\quad t\in[0,\pi/2\Omega]; \nonumber\\
&& \Omega_{1}=4\Omega/\sqrt{6},\quad \Omega_{0,2,3}=0,\quad t\in(\pi/2\Omega,\pi/\Omega]; \nonumber\\
&& \Omega_{2}=4\Omega/\sqrt{6},\quad \Omega_{0,1,3}=0,\quad t\in(\pi/\Omega,3\pi/2\Omega]; \nonumber\\
&& \Omega_{3}=2\Omega,\quad \Omega_{0,1,2}=0,\quad t\in(3\pi/2\Omega,2\pi/\Omega].
\end{eqnarray}
In this situation, the particle hopping process in the five-site DSL is divided into four times of 1-link hypercube, and at the end of each step the particle fully reaches a certain site, which shows the searching ability of DSL for finding any one site.

\begin{figure}[t]
	\centering
	\includegraphics[width=0.96\linewidth]{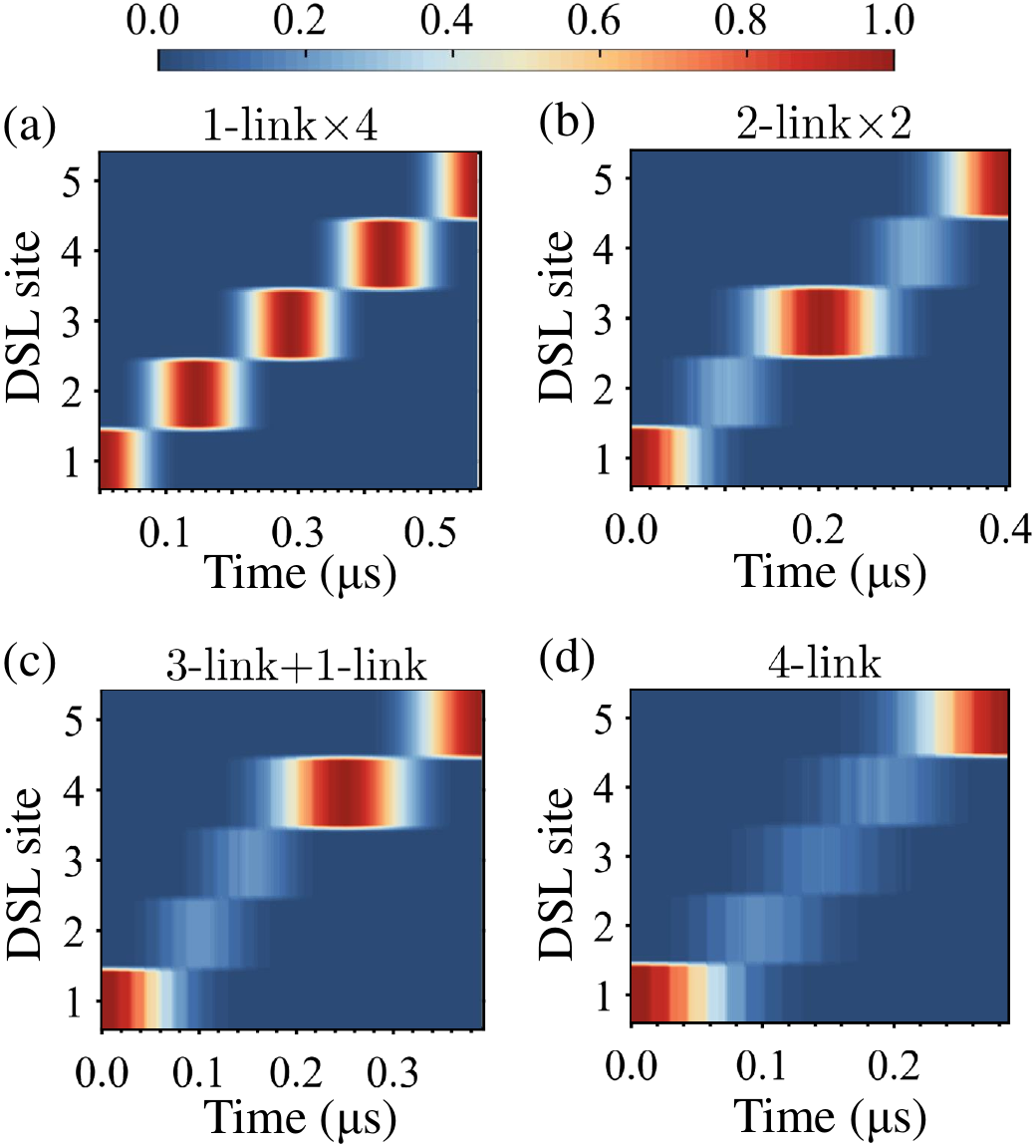}
	\caption{\label{fig3} Full RAB processes through population transfers from $|2,-2\rangle$ to $|2,2\rangle$ in different ways based on the PST method. (a)~Four steps via four times of 1-link hops. (b)~Two steps via two times of 2-link hops. (c)~Two steps via one 3-link hop plus one 1-link hop. (d)~One step via one 4-link hop. Detailed parameters and descriptions are found in the main text.}
\end{figure}
In addition to the nearest-neighbor 1-link hypercube, the long-range state transfer is also accessible in DSL. In Figure~\ref{fig3}(b), the particle hops via a 2-link$\times$2 process to the fifth site by two steps with one time of quenched Hamiltonian, with parameters
\begin{eqnarray}
&& \Omega_{0}=\Omega,~\Omega_{1}=4\Omega/\sqrt{6},~\Omega_{2,3}=0,~t\in[0,\pi/\sqrt{2}\Omega]; \\
&& \Omega_{2}=4\Omega/\sqrt{6},~\Omega_{3}=2\Omega,~\Omega_{0,1}=0,~ t\in(\pi/\sqrt{2}\Omega,\sqrt{2}\pi/\Omega].\nonumber
\end{eqnarray}
In Figure~\ref{fig3}(c), the particle hops via a 3-link$+$1-link process to the fifth site with
\begin{eqnarray}
&& \Omega_{0}=\Omega,\quad\Omega_{1}=8\Omega/3\sqrt{2},\quad\Omega_{2}=4\Omega/\sqrt{6},\nonumber\\
&&\Omega_{3}=0,\quad t\in[0,\sqrt{3}\pi/2\Omega];\\
&& \Omega_{3}=2\Omega,\quad \Omega_{0,1,2}=0,\quad t\in(\sqrt{3}\pi/2\Omega,(\sqrt{3}+1)\pi/2\Omega]. \nonumber
\end{eqnarray}
In addition to these stepwise ways, the particle can also hop to the fifth site in a single step through a direct 4-link process without quenched Hamiltonian, as shown in Figure~\ref{fig3}(d), with
\begin{equation}
\Omega_{0}=\Omega,~\Omega_{1}=2\Omega,~\Omega_{2}=2\Omega,~\Omega_{3}=2\Omega,~t\in[0,\pi/\Omega].
\end{equation}
The processes above exhibit the capacity of tunable, flexible and programmable quantum searching and long-range state transfers in our DSL, and it enables the generation of the four-atom entangled states: Twin-Fock state $|2,0\rangle$; $W$ states $|2,-1\rangle$ and $|2,1\rangle$.

	It is worth noting that a sequential, atom-by-atom spin-flip excitation (e.g., $|0000\rangle \rightarrow |1000\rangle \rightarrow \dots \rightarrow |1111\rangle$) can also be achieved in principle via a stepwise antiblockade mechanism \cite{SLSu2017PRA_fast}. Such stepwise schemes typically require applying distinct laser detunings to individual atoms to satisfy the resonance conditions sequentially. However, this sequential approach necessitates multi-step pulse operations, demands high timing precision for each step, and incurs a total operation time that scales linearly with the number of atoms. In contrast, the scheme presented here preserves the permutation symmetry of the atoms and utilizes Floquet modulation to generate a synthetic dimension. By employing a single global driving field containing multiple harmonic components, all adjacent transitions within DSL are activated simultaneously. This single-step continuous driving dramatically reduces the total time cost and significantly simplifies the required experimental control compared to sequential stepwise operations.\\

		\begin{figure}[t]
		\centering
		\includegraphics[width=\linewidth]{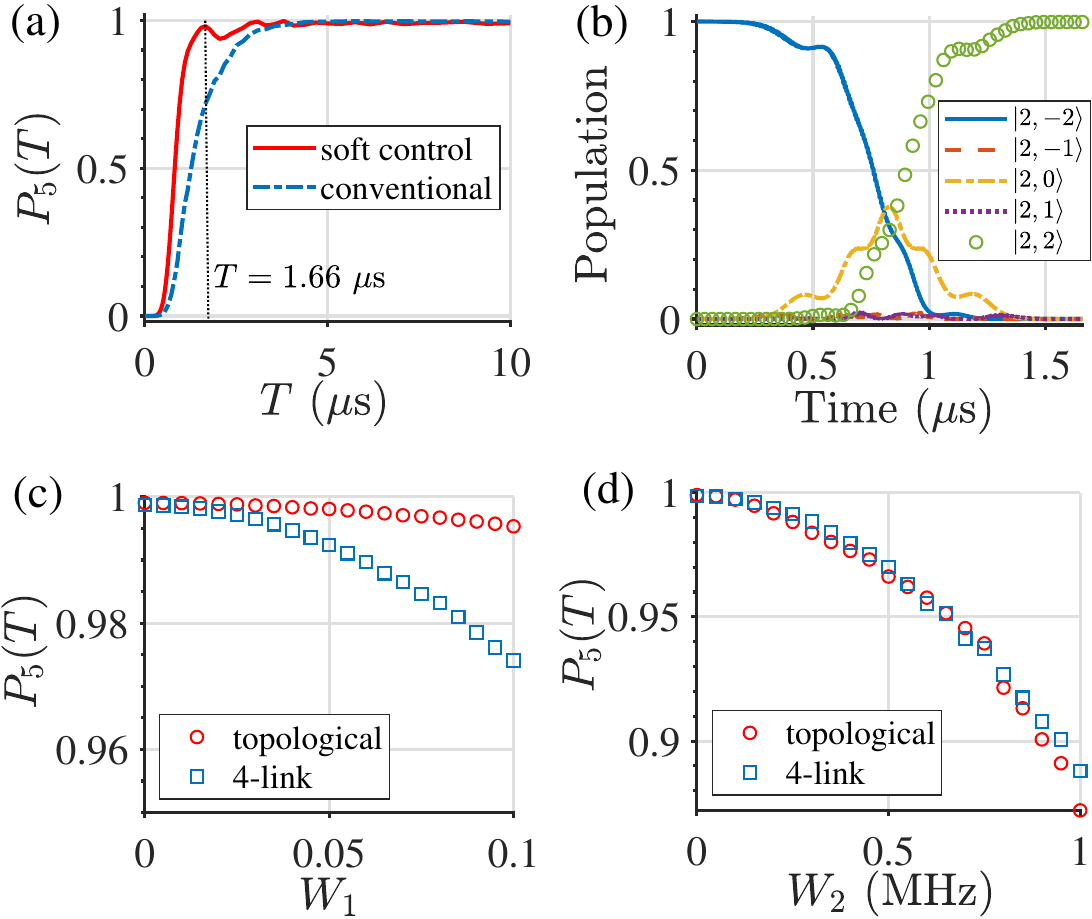}
		\caption{\label{fig4}(a) Population of $|1111\rangle$ at the instant $t = T$ versus the duration $T$, based on the conventional scheme and the soft control scheme with the Gaussian envelopes in eq.~\eqref{eq11}. (b) Time-dependent populations of the five Dicke states during the transition from the ground state to the full RAB state. Effect of disorder intensity on the final population of $|1111\rangle$ for (c) off-diagonal terms and (d) diagonal terms, with comparison between the soft control scheme and the 4-link PST scheme. Each datum in (c) and (d) is obtained by averaging 501 results with the off-diagonal or diagonal disorder terms being randomly picked.}
	\end{figure}
	\noindent \textbf{\begin{large}4~~{Simulation of topological state transfer}\end{large}}\\
	\\
	We now show the capacity of DSL in quantum simulation, for which we implement topological edge-to-edge state transfer in a small SSH lattice~\cite{LQi2020OL,FMei2018PRA,JKGuo2024PRA}, through a collective spin flip $|0000\rangle\rightarrow|1111\rangle$ by the advantage of soft quantum control~\cite{Haase2018PRL}. We construct the SSH model by setting the intercell hopping rate $\mathcal{V}(t)=\Omega_0(t)=\frac{\sqrt{6}\Omega_2(t)}{4}$ and the intracell hopping rate $\mathcal{W}(t)=\frac{\sqrt{6}\Omega_1(t)}{4}=\frac{\Omega_3(t)}{2}$. The conventional envelopes with temporal modulation of hopping hold trigonometric types $\mathcal{V}(t)=\Omega\sin^2(\pi t/2T)$ and $\mathcal{W}(t)=\Omega\cos^2(\pi t/2T)$~\cite{LQi2020OL,FMei2018PRA},
	with $T$ being the total duration. To improve the performance of topological state transfer under rotating-wave approximation (suppressing off-resonant terms), we can employ soft quantum control~\cite{Haase2018PRL,HDYin2021OL,XXLi2021PRApplied,QWu2024EPJQT,JKGuo2024PRA,JWu2025npjQI,JLWu2022SCPMA} with Gaussian temporal envelopes of $\mathcal{V}(t)$ and $\mathcal{W}(t)$
	\begin{equation}\label{eq11}
	\mathcal{V}(t)=\Omega e ^{-(t-T/2-\tau_d)^2/\sigma^2},\quad \mathcal{W}(t)=\Omega e ^{-(t-T/2+\tau_d)^2/\sigma^2},
	\end{equation}
	with $\tau_d = T/10$ and $\sigma = T/4$. Due to starting from and ending at zero amplitudes, the Gaussian envelopes can more exactly select resonant transitions in the system and simultaneously suppress unwanted contributions coming from off-resonant transitions~(see ref.~\cite{Haase2018PRL} for more details).
	More clearly, we show the final population of the full RAB state $|1111\rangle$ with varying duration by specifying $\Omega = 2\pi\times 1.75$ MHz in Figure~\ref{fig4}(a). This comparison demonstrates the advantage of soft quantum control with Gaussian envelopes, where the full excitation population $P_5(T)$ reaches 0.99 with $T=1.66~\mu$s when the conventional scheme is $P_5(T)\sim0.75$. In Figure~\ref{fig4}(b), we plot the time-dependent populations of five DSL sites under Gaussian soft quantum control with $T=1.66~\mu$s, exhibiting a complete topological state transfer process of a collective spin flip $|0000\rangle\rightarrow|1111\rangle$.
	
	Subsequently, in Figures~\ref{fig4}(c) and \ref{fig4}(d), we investigate effects of off-diagonal and diagonal disorder, respectively, on the fidelity, i.e., $P_5(T)$, of achieving the collective spin flip based on the soft control scheme and the 4-link PST scheme. We introduce the off-diagonal disorder intensity $W_1$ by $\Omega_n(t)\to\Omega_n(t)(1+\delta_1^n)$ with randomly picked $\delta_1^n\in[-W_1,W_1]$, and the diagonal disorder intensity $W_2$ by $\hat{H}(t)\to \hat{H}(t)+\sum_{l=1}^5\delta_2^l|2,l-3\rangle\langle 2,l-3|$ with randomly picked $\delta_2^l\in[-W_2,W_2]$.
	We observe that the 4-link PST scheme is largely affected by the off-diagonal disorder, with the final population $P_5(T)$ of the DSL site $|2,2\rangle$ declining significantly when increasing $W_1$, which can be attributed to the fundamental dependence of the PST scheme on the effective pulse area~\cite{JLWu2025NC}. Regarding the off-diagonal disorder intensity in Figure~\ref{fig4}(c), it demonstrates remarkable robustness against disturbances when the topological state transfer scheme with the soft quantum control is applied. This resilience stems from its independence on the pulse area, and moderate deviations in off-diagonal parameters do not break the chiral symmetry in the SSH model~\cite{LQi2020OL,FMei2018PRA}. However, diagonal disorder breaks chiral symmetry and undermines the topological protection of the SSH system. As shown in Figure~\ref{fig4}(d), the final populations $P_5(T)$ in both the 4-link PST scheme and the topological state transfer scheme significantly decrease with the same magnitude of reduction when increasing the diagonal disorder intensity $W_2$. The emergence of these typical characteristics belonging to the standard PST and topological state transfer sufficiently proves the strong capacity of our cascaded RAB in exploring collective spin excitation dynamics and simulating topological dynamics behaviors.
	
	Figures \ref{fig5}(a) and \ref{fig5}(b) show the instantaneous energy spectra of the five-site DSL under the conventional modulation scheme and the Gaussian soft-control scheme, respectively. These figures plot the variation of the instantaneous eigenenergies $E_n(t)$ as a function of the normalized time $t/T$. Furthermore, Figures \ref{fig5}(c)–\ref{fig5}(e) display the spatial probability distributions of the gap state across the five lattice sites at three distinct instants: $t = T/4$, $T/2$, and $3T/4$. It can be clearly observed that at $t = T/4$, the gap state is primarily localized at the leftmost site ($|2,-2\rangle$). At $t = T/2$, the wave function delocalizes over the entire chain, exhibiting bulk-state characteristics. Finally, at $t = 3T/4$, the gap state successfully transfers to the rightmost site ($|2,2\rangle$). This time-resolved evolution intuitively verifies the adiabatic transfer of the topological edge state from the left side to the right side, which perfectly aligns with the high-fidelity full-excitation results achieved under the Gaussian soft-control scheme shown in Figure 4(b).\\
	\begin{figure}[t]
		\centering
		\includegraphics[width=\linewidth]{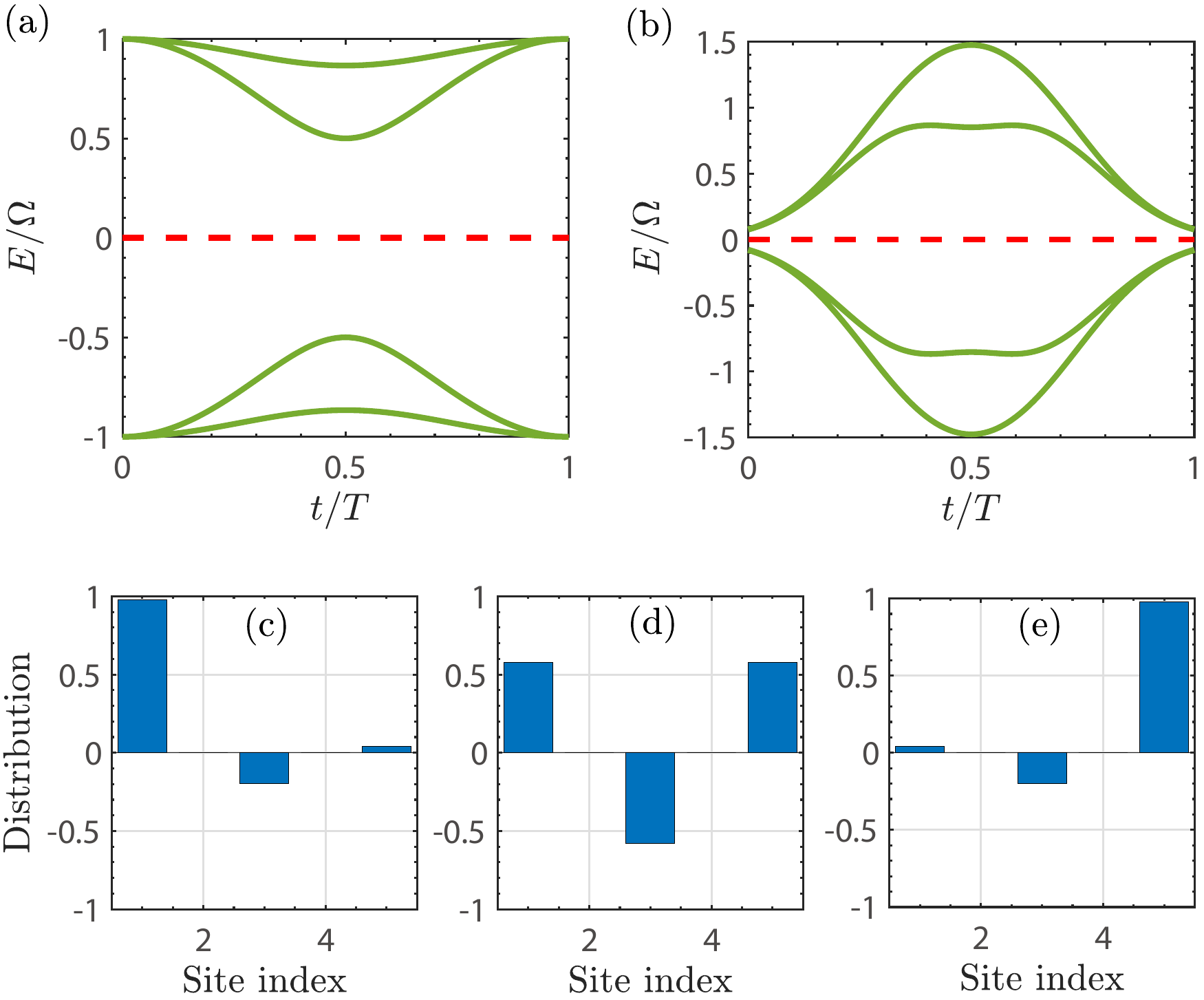}
		\caption{Instantaneous energy spectrum of the five-site DSL under (a) the conventional modulation and (b) the Gaussian soft-control scheme. (c)–(e) Spatial distributions of the gap state on the five sites at $t = T/4$, $T/2$, and $3T/4$ for the Gaussian scheme.}\label{fig5}
	\end{figure}
	\\
	
	\begin{figure}[t]
		\centering
		\includegraphics[width=\linewidth]{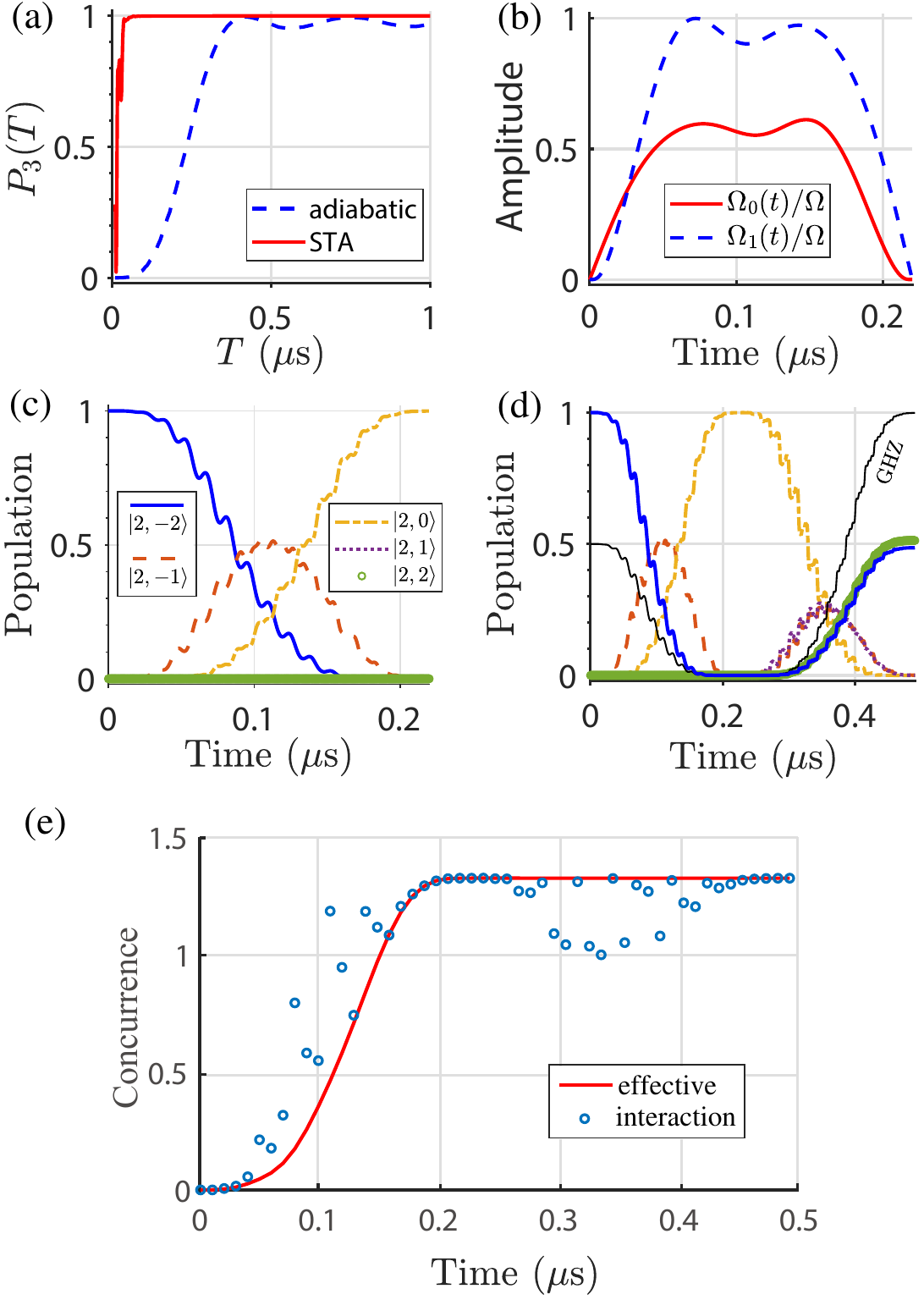}
		\caption{\label{fig6}(a) Population of $|2,0\rangle$ at the instant $t = T$ versus the duration $T$, based on the adiabatic scheme with the Gaussian envelopes  in eq.~\eqref{eq11} and the STA scheme with $\gamma = 0.25\pi$. (b) Temporal envelopes of STA $\Omega_{0,1}(t)$. Time-dependent populations of the five Dicke states during the transition from the fully ground state $|2,-2\rangle=|0000\rangle$ to (c) the twin-Fock state and (d) the generated GHZ state, respectively. (e) Schematic diagram using concurrence to quantitatively characterize the entanglement properties of the prepared twin-Fock state and GHZ state. The red solid line represents the effective Hamiltonian, while the blue circle represents the interaction Hamiltonian in eq.~\eqref{eq1}.}
	\end{figure}	
	\noindent \textbf{\begin{large}5~~~{Generating entanglement via STA}\end{large}}\\
	\\
	Finally, we identify the flexibility of our cascaded RAB with DSL in incorporating existing advanced control methods for generating multi-atom entanglement~\cite{RHZheng2020PRA,Haase2021PRA,Haase2022PRResearch,Stojanovic2022PRA,Nauth2022PRA,XQShao2023PRA,Ramaswamy2025PRResearch,YZhao2024APL}. Concretely, when reducing the number of components of the drive laser to two, that is, $\Omega_2(t)=\Omega_3(t)=0$, the effective SSH Hamiltonian becomes
	\begin{equation}\label{eq13}
	\hat{H}_{\mathrm{eff}}^{\prime}(t)=\mathcal{V}(t)|2,-2\rangle\langle 2,-1|+\mathcal{W}(t)|2,-1\rangle\langle 2,0|+\mathrm{H.c.},
	\end{equation}
	which involves only three-state transitions $|2,-2\rangle\leftrightarrow|2,-1\rangle\leftrightarrow|2,0\rangle$. It is convenient to generate the twin-Fock entangled state through the technique of STA.
	
	Various pulse-shaping methods have been developed for STA implementation~(see, e.g., the review article~\cite{STA2019RMP}). Here we refer to our previous STA-related works in three-level systems~\cite{JLWu2017OE,XKSong2016PRA,JLWu2019PRA}, and set (see Appendix~C\ref{Appendix_C} for details)
	\begin{eqnarray}\label{eq14}
	\mathcal{V}(t)=\dot{\theta}(t)\sin\theta(t)\cot\mu(t)+\dot{\mu}(t)\cos\theta(t),\nonumber\\ \mathcal{W}(t)=\dot{\theta}(t)\cos\theta(t)\cot\mu(t)-\dot{\mu}(t)\sin\theta(t),
	\end{eqnarray}
	where the involved parameters are
	\begin{eqnarray}\label{e14}
	&&\mu=\gamma\sin^{2}\frac{\pi t}{T}+\epsilon,\quad\dot{\mu}=\frac{\gamma\pi}{T}\sin\frac{2\pi t}{T},\nonumber\\
	&&\theta=\sum_{j=5}^{9}a_{j}t^{j},\quad\dot{\theta}=\sum_{j=5}^{9}ja_{j}t^{j-1},\nonumber\\
	&&a_{5}=\frac{63\pi}{T^{5}},\quad a_{6}=-\frac{210\pi}{T^{6}},\quad a_{7}=\frac{270\pi}{T^{7}},\nonumber\\
	&&a_{8}=-\frac{315\pi}{2T^{8}},\quad a_{9}=\frac{35\pi}{T^{9}}.
	\end{eqnarray}
$\epsilon = 0.01$ is a small value to guarantee finite $\mathcal{V}(t)$ and $\mathcal{W}(t)$.	In Figure~\ref{fig6}(a) we compare the adiabatic scheme based on the Gaussian envelopes in eq.~\eqref{eq11} and the STA scheme with $\gamma = 0.25\pi$ by plotting the population $P_3(T)$ of the twin-Fock entangled state $|2,0\rangle$ versus the duration $T$. The STA scheme significantly outperforms for generating the twin-Fock entangled state than the adiabatic scheme. Specifically, with $T = 0.22~\mu$s being assigned to ensure $\max[\Omega_{0,1}(t)]=\Omega=2\pi\times1.75$~MHz, the STA generation of the twin-Fock state is completed, while the adiabatic generation is almost entirely ineffective. The temporal envelopes of $\Omega_{0}(t)$ and $\Omega_{1}(t)$ , considering $\Omega_{2}(t) = \Omega_{3}(t) = 0$, are shown in Figure~\ref{fig6}(b), and the time-dependent populations of the five Dicke states are shown in Figure~\ref{fig6}(c).
	
	For generating GHZ state, we set the relations $\frac{\sqrt{3}\Omega_1(t)}{2} = \frac{\sqrt{3}\Omega_2(t)}{2} = \mathcal{V}(t)$ and $\Omega_0(t) = \frac{\Omega_3(t)}{2} = \mathcal{W}(t)$, and the effective Hamiltonian in eq.~\eqref{eq3} becomes
	\begin{equation}
	\hat{H}_{\mathrm{eff}}^{\prime\prime}=\mathcal{V}(t)|2,0\rangle\langle W|+\mathcal{W}(t)|W\rangle\langle\mathrm{GHZ}|+\mathrm{H.c.},
	\end{equation}
	where $|W\rangle=(|2,-1\rangle+|2,1\rangle)/\sqrt{2}$ and $|\mathrm{GHZ}\rangle=(|0000\rangle+|1111\rangle)/\sqrt{2}$. We then can use the STA scheme, similar to the process of generating the twin-Fock state $|2,0\rangle$, to transfer $|2,0\rangle$ to the GHZ state of four atoms. To complete the generation of GHZ state, we connect the process of generating the twin-Fock state from the fully ground state $|2,-2\rangle$ (setting the duration $T_1 = 0.22~\mu$s) to the process of transferring $|2,0\rangle$ to the GHZ state (setting the duration $T_2 = 0.27~\mu$s), with $|0000\rangle$ being the initial state. The time-dependent populations of the five Dicke states and the GHZ state are shown in Figure~\ref{fig6}(d), which shows a complete procedure for generating a high fidelity GHZ state of the four atoms.
	
	To quantitatively characterize the entanglement properties of the prepared twin-Fock state and GHZ state, we adopt the generalized concurrence as a genuine multipartite entanglement measure, which is defined as~\cite{PhysRevLett.93.230501,PhysRevLett.95.260502}
			\begin{equation}\label{e16}
			C_N(\Psi) = 2^{1-N/2} \sqrt{ (2^N-2) - \sum_{\alpha} \operatorname{Tr}(\rho_\alpha^2) },
			\end{equation}
			where $N$ represents the number of particles and $\rho_\alpha$ denotes the reduced density matrix of the subsystem $\alpha$. The summation runs over all $2^N-2$ possible proper subsets of the $N$-particle system. Figure~\ref{fig6}(e) illustrates the time evolution of the concurrence during the state transfer process under the STA scheme. The results indicate that at the optimal preparation time ($t=0.27~\mu\text{s}$), the concurrence of the twin-Fock state reaches $1.32$, which is close to the analytical theoretical maximum ($\sqrt{7}/2 \approx 1.3228$), confirming that the generated four-atom twin-Fock state possesses extremely high entanglement quality. Meanwhile, for the cascaded full excitation, the concurrence of the generated GHZ state also reaches $1.32$ at $t = 0.49~\mu\text{s}$. This quantitative analysis not only rigorously verifies the high operational fidelity of the STA scheme but also provides a solid foundation for utilizing these dynamically generated states as reliable resources for multipartite quantum information processing.\\
	\\
	\noindent \textbf{\begin{large}6~~~{Discussion and conclusion}\end{large}}\\
	\\
	\begin{figure}[t]
		\centering
		\includegraphics[width=0.8\linewidth]{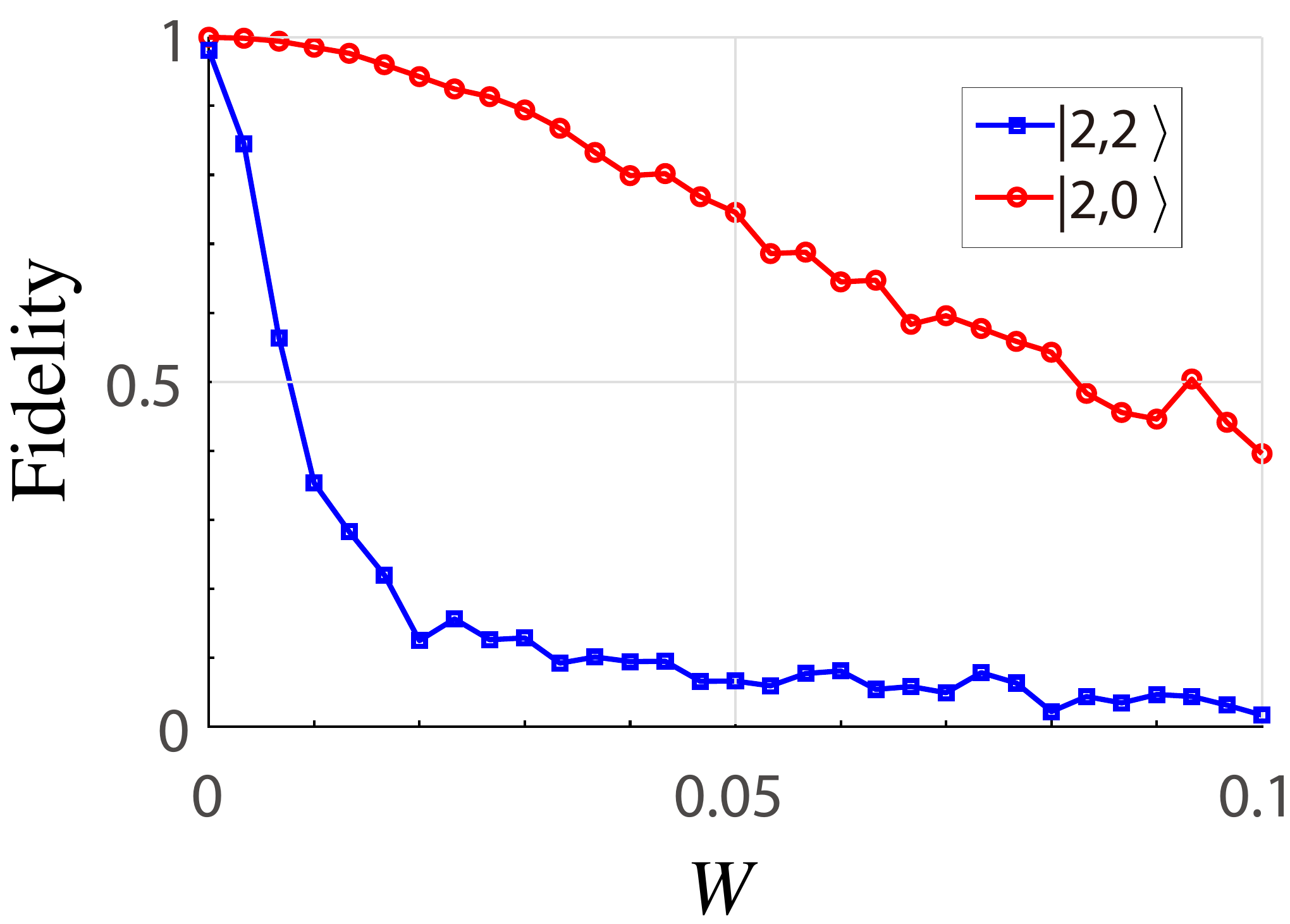}
		\caption{Effect of the relative vdW interaction error strength $W$ on the entanglement fidelity of the twin-Fock state $|2,0\rangle$ and the fully excited state $|2,2\rangle$. Each data point represents the average of 501 independent numerical simulations with $V_{jk}=\delta V_{jk}+V$ by randomly sampling independent four $\delta V_{jk}$ from $\delta V_{jk}/V\in[-W, W]$.}
		\label{fig7}
	\end{figure}
It is essential to characterize the robustness of our protocol against trap-induced position uncertainty and finite temperature effects. Due to the strict vdW scaling ($V \propto 1/d^6$), position fluctuations $\delta d$ result in an amplified interaction error $\delta V/V \approx -6 \delta d/d$. As shown in Figure~\ref{fig7}, we numerically simulate the effect of the relative interaction error strength $W$ on state transfer, which results in relative vdW interaction deviations $\delta V_{jk}/V\in[-W,~W]$ from the desired value $V$ for the four-atom configuration, with $\delta V_{jk}=V_{jk}-V$. For atoms cooled to the motional ground state in optical tweezers ($\delta d \sim 30$ nm~\cite{Levin2018PRL}), the relative position uncertainty at $d = 4.8~\mu$m is approximately $0.6\%$, corresponding to a realistic relative interaction error strength $W \approx 3.6\%$. Under this experimental condition, the system reveals a compelling dichotomy. On one hand, the twin-Fock state $|2,0\rangle$ maintains a robust fidelity of $>0.8$, confirming the experimental feasibility of the Dicke-state lattice for generating specific multipartite entangled states. On the other hand, the fully excited state $|1111\rangle$ drops sharply in fidelity to below $0.1$. This extreme sensitivity demonstrates that while the full cascaded RAB process is challenging to sustain for a high-fidelity state transfer under finite temperature effects, it can be leveraged as a unique resource for quantum precision measurement. By setting the driving frequency $\omega$ exactly resonant with $V$, the system acts as a highly sensitive multiphoton probe. Any minute perturbation altering Rydberg states, such as ambient stray electric fields causing AC Stark shifts or spatial drifts, will dramatically suppress the $|1111\rangle$ population. Inspired by the recently reported criticality-enhanced metrology~\cite{DSDing2022NP}, our highly programmable platform holds promising potential as a high-resolution quantum sensor, turning the vulnerability of Rydberg interactions into a metrological advantage.\\

Furthermore, it is necessary to explicitly position our protocol within the broader context of multipartite state transfer and entanglement generation. In recent years, numerous excellent theoretical schemes have addressed the generation and interconversion of $W$ and GHZ states in strongly interacting systems. Because generating a GHZ state intrinsically requires populating fully excited components (e.g., $|111\rangle$), achieving this in the blockade regime effectively necessitates overcoming the interaction penalty, functioning as an effective antiblockade. Existing works typically navigate this strongly interacting manifold by employing sophisticated algorithmic optimal control, such as Lewis-Riesenfeld invariants \cite{RHZheng2020PRA}, Lie-algebraic parameterizations \cite{Haase2021PRA}, quantum-brachistochrone formalisms \cite{Nauth2022PRA}, customized multi-pulse sequences \cite{Stojanovic2022PRA}, or dissipative Floquet-Lindblad engineering \cite{XQShao2023PRA}. Furthermore, related entanglement transfers have been explored using generalized $W$ states \cite{Haase2022PRResearch} and atom-cavity interactions \cite{Ramaswamy2025PRResearch}.  The essential novelty of our work, distinguishing it from the aforementioned state-specific optimization protocols, resides in the synthetic-dimension interpretation. By applying a global Floquet amplitude modulation, we do not merely force a specific state-to-state conversion. Instead, we explicitly construct a resonant, first-order cascaded RAB mechanism. This physically maps the complex many-body excitation dynamics into a highly controllable single-particle hopping model within a 1D synthetic dimension (DSL). Consequently, sophisticated multi-atom entanglement generation (e.g., twin-Fock and GHZ states) becomes a natural and mathematically straightforward application of single-particle control techniques (like STA and topological SSH edge state transfer) within this unified, globally driven platform.
	
		Our work establishes a flexible and programmable synthetic dimension for engineering multipartite entanglement and simulating many-body dynamics in Rydberg atom arrays. Looking forward, the cascaded RAB and DSL framework can be extended to explore richer physical phenomena. For instance, introducing onsite potentials may lead to the realization of Wannier-Stark ladders and the study of localization dynamics in excitation-number space~\cite{XYGuo2021NPJQI,RMao2022Light,TChen2024PRL,SZLi2025SCPMA}. Incorporating next-nearest-neighbor and long-range hoppings into the DSL could reveal novel interference effects~\cite{TChen2024NC,TChen2025NP}. Furthermore, combining STA with topological state transfer may enable faster and more robust edge-state transfers and collective spin flips~\cite{JKGuo2024PRA,JLWu2025NC}. These directions highlight the versatility of our approach and its promise for advancing multiparticle quantum information processing with neutral atoms.

	In conclusion, we have proposed and systematically investigated a cascaded RAB regime in four fully connected interacting atoms via Floquet modulation. By constructing a synthetic DSL in the space of collective spin excitations, we have demonstrated PST across the five-site lattice with multiple programmable pathways, ranging from stepwise nearest-neighbor hopping to direct single-step transitions. This DSL platform further enables quantum simulation of topological dynamics, where soft quantum control is employed to achieve topologically inspired full RAB $|0000\rangle \to |1111\rangle$ with enhanced robustness against disorder. Moreover, by incorporating STA techniques, we have realized high-fidelity generation of twin-Fock and GHZ entangled states on the four atoms within sub-microsecond timescales, overcoming the speed limitations of conventional adiabatic protocols.\\
	
		\begin{small}\textit{The authors acknowledge funding from the  National Natural Science Foundation of China (NSFC) (12304407, 62571494, 12575032, 12274376, 62471001); Natural Science Foundation of Henan Province (262300421244, 262300421245, 262300422574, 232300421075); China Postdoctoral Science Foundation~(2023TQ0310, GZC20232446, 2024M762973); The Open Project of the State Key Laboratory of Metabolic Dysregulation \& Prevention and Treatment of Esophageal Cancer (2025SGAQZ-QN-05).}\end{small}\\

\noindent \textbf{Conflict of interest\quad}The authors declare that they have no conflict of interest.
	\maketitle
	
	\bibliography{apssamp}


	\appendix
	\begin{widetext}
\section{Appendix A~~~Derivation of $\hat{H}_{\mathrm{eff}}(t)$}\label{Appendix_A}
Based on eq.~\eqref{eq1} and eq.~\eqref{eq2}, the Hamiltonian of the four-Rydberg-atom system under Floquet modulation can be written as
\begin{equation}
\hat{H}_1(t)=\sum_{j=1}^{4}\sum_{n=0}^{3}\frac{\Omega_{n}(t)}{4}\left(e^{- i nwt}+e^{ i nwt}\right)\hat{\sigma}_{x}^{j}+\sum_{j>k=1}^{4}V_{jk}|11\rangle_{jk}\langle11|,
\end{equation}
Because the four spins are identical with respect to the global driving and RRI, the Hamiltonian can be expressed by the four-spin Dicke states in eq.~\eqref{eq4} that are eigenstates of the RRI Hamiltonian $\sum_{j>k=1}^{4}V_{jk}|11\rangle_{jk}\langle11|$ with eigenenergies in eq.~\eqref{eq5}.
In the Dicke-state space by expanding the Floquet-modulated pulse, the total Hamiltonian becomes
	\begin{eqnarray}
\hat{H}_2(t)	&=&\frac{1}{2}\left[2\Omega_{0}(t)+\Omega_{1}(t)\left(e^{- i wt}+e^{ i wt}\right)+\Omega_{2}(t)\left(e^{- i 2wt}+e^{ i 2wt}\right)+\Omega_{3}(t)\left(e^{- i 3wt}+e^{ i 3wt}\right)\right]|2,-2\rangle\langle2,-1| \nonumber\\
	&&+\frac{\sqrt{6}}{4}\left[2\Omega_{0}(t)+\Omega_{1}(t)\left(e^{- i wt}+e^{ i wt}\right)+\Omega_{2}(t)\left(e^{- i 2wt}+e^{ i 2wt}\right)+\Omega_{3}(t)\left(e^{- i 3wt}+e^{ i 3wt}\right)\right]e^{- i Vt}|2,-1\rangle\langle2,0| \nonumber\\
	&&+{\frac{\sqrt{6}}{4}}\left[2\Omega_{0}(t)+\Omega_{1}(t)\left(e^{- i wt}+e^{ i wt}\right)+\Omega_{2}(t)\left(e^{- i 2wt}+e^{ i 2wt}\right)+\Omega_{3}(t)\left(e^{- i 3wt}+e^{ i 3wt}\right)\right]e^{- i 2Vt}|2,0\rangle\langle2,1| \nonumber\\
	&&+{\frac{1}{2}}\left[2\Omega_{0}(t)+\Omega_{1}(t)\left(e^{- i wt}+e^{ i wt}\right)+\Omega_{2}(t)\left(e^{- i 2wt}+e^{ i 2wt}\right)+\Omega_{3}(t)\left(e^{- i 3wt}+e^{ i 3wt}\right)\right]e^{- i 3Vt}|2,1\rangle\langle2,2| \nonumber\\
	&&+\mathrm{H.c.}
	\end{eqnarray}
Under both the RAB  condition and the rotating-wave approximation (RWA) condition $V=\omega\gg\mathrm{max}[\Omega_{n}(t)]$, we neglect the off-resonant terms and finally obtain an effective Hamiltonian
	\begin{equation}
	\hat{H}_{\mathrm{eff}}(t)=\Omega_{0}(t)|2,-2\rangle\langle2,-1|+\frac{\sqrt{6}\Omega_{1}(t)}{4}|2,-1\rangle\langle2,0|+\frac{\sqrt{6}\Omega_{2}(t)}{4}|2,0\rangle\langle2,1|+\frac{\Omega_{3}(t)}{2}|2,1\rangle\langle2,2|+\mathrm{H.c.}
	\end{equation}
	
\section{Appendix B~~~Two-photon Rydberg pumping}
In order to implement the transition from a ground state to a Rydberg
state, we consider a two-photon process in $^{87}$Rb atoms~\cite{Evered2023Nature,Levin2018PRL,Levin2019PRL}. The related energy levels of $^{87}$Rb atoms are set as $|0\rangle\equiv|5S_{1/2}, F = 1, m_{F}=0\rangle$ and $|1\rangle\equiv|70S_{1/2}, m_{J}=-1/2\rangle$ with the vdW interaction coefficient $C_6/2\pi=858.4~{\rm GHz}~\mu {\rm m}^6$. An intermediate state $|p\rangle=|5p_{3/2}\rangle$ is specified to mediate this two-photon transition achieved by imposing two laser fields. One field works for the optical excitation $|0\rangle\leftrightarrow|p\rangle$ with Rabi frequency $\Omega_{0p}$ and a blue detuning $\Delta_{p}$, while the other is imposed for the Rydberg excitation $|p\rangle\leftrightarrow|1\rangle$ with Rabi frequency $\Omega_{1p}$ and a red detuning $\Delta_{p}$. The Floquet modulation is implemented by the field for the transition $|p\rangle\leftrightarrow|1\rangle$, that is, $\Omega_{1p}(t)=\sum_{n=0}^{N}\Omega_{1p}^{(n)}(t)\cos(n\omega t)$.

 $\Delta_p$ is so large that the intermediate state $|p\rangle$ can be adiabatically eliminated and the two-photon resonant Rydberg pumping from the ground state $|0\rangle$ to the Rydberg state $|1\rangle$ can be achieved. The resulting effective Rabi frequency of the wanted transition $|0\rangle\leftrightarrow|1\rangle$ is Floquet modulated
 \begin{equation}
 \Omega_{01}(t)=\frac{\Omega_{0p}\Omega_{1p}(t)}{2\Delta_p}=\sum_{n=0}^N \frac{\Omega_{0p}\Omega_{1p}^{(n)}(t)}{2\Delta_p}\cos(n\omega t).
 \end{equation}
When we set $\Omega_n(t)={\Omega_{0p}\Omega_{1p}^{(n)}(t)}/{2\Delta_p}$, the effective Rabi frequency $\Omega_{01}(t)$ is exactly the desired one in eq.~\eqref{eq2}. In our work, the magnitude of amplitude components in the desired $\Omega(t)/2\pi$ is several MHz, which can be achieved by setting experimentally feasible parameters $\Delta_p/2\pi\sim1$~GHz, $\Omega_{0p}/2\pi\sim100$~MHz, and $\max[\Omega_{1p}^{(n)}(t)]/2\pi\sim50$~MHz.

In addition to the desired two-photon effective Rydberg pumping, it is noted that there are Stark shifts $\Omega_{0p}^2/4\Delta_{p}|0\rangle\langle0|$ for the ground state and $\Omega_{1p}^2(t)/4\Delta_{p}|1\rangle\langle 1|$ for the Rydberg state. To counteract these unwanted energy shifts, one can either employ additional lasers to drive off-resonant transitions to auxiliary states, generating opposite Stark shifts~\cite{JLWu2022FOP}, or implement phase corrections~\cite{Vepsalainen2019SA}.

\section{Appendix C~~~STA pulse engineering}\label{Appendix_C}
For achieving STA in the equivalent three-level system described by the Hamiltonian in eq.~\eqref{eq13}, we seek a dressed state that serves as the evolutionary path, ensuring that the initial state is $|2,-2\rangle$ and the final state is $|2,0\rangle$. This state can be chosen as~\cite{JLWu2017OE}
	\begin{equation}
	|\Psi_0(t)\rangle=\cos\mu(t)[\cos\theta(t)|2,-2\rangle+\sin\theta(t)|2,0\rangle]+e^{i\phi}\sin\mu(t)|2,-1\rangle,
	\end{equation}
	where $\mu(t)$ is an auxiliary parameter. $\theta(t)$ is defined by the mixed angle $\theta(t)=\arctan[\mathcal{V}(t)/\mathcal{W}(t)]$, which yields $\mathcal{V}(t)=\mathcal{U}(t)\sin\theta(t)$ and $\mathcal{W}(t)=\mathcal{U}(t)\cos\theta(t)$ with $\mathcal{U}(t)=\sqrt{\mathcal{V}^2(t)+\mathcal{W}^2(t)}$. As long as $|\Psi_0(t)\rangle$ serves as the evolutionary path with the introduced parameter $\mu(t)$ satisfying the boundary conditions $\mu(0) = \mu(t) = 0$, it is evident that the desired state can be obtained. To form a complete set of orthogonal basis states, the other two dressed states $|\Psi_{\pm}(t)\rangle$ must satisfy $\langle \Psi_i(t) | \Psi_j(t) \rangle = \delta_{ij}$ (where $i,j = \pm, 0$; $\delta_{ij} = 1$ if $i = j$, and $\delta_{ij} = 0$ otherwise) as well as the completeness condition $\sum_{n=\pm,0}|\Psi_n(t)\rangle\langle\Psi_n(t)|=1$. These two dressed states can be chosen as
\begin{equation}
|\Psi_{\pm}(t)\rangle
=\quad\frac{1}{\sqrt{2}}\{[\sin\theta(t)\mp i\sin\mu(t)\cos\theta(t)]|2,-2\rangle\mp\cos\mu(t)|2,-1\rangle-[\cos\theta(t)\pm i\sin\mu(t)\sin\theta(t)]|2,0\rangle\},
\end{equation}
where, for simplicity, we have set $\phi=\pi/2$.

To ensure that $|\Psi_0(t)\rangle$ serves as the evolutionary path of the system, it is necessary to carefully design pulse envelope such that the transitions between dressed states $|\Psi_0(t)\rangle$ and $|\Psi_{\pm}(t)\rangle$ are eliminated. To this end, the original Rabi frequencies $\mathcal{V}(t)$ and $\mathcal{W}(t)$ can be modified by  $\mathcal{V}(t)\to \mathcal{U}(t)\sin\theta(t)+\mathcal{V}'(t)$ and $\mathcal{W}(t)\to \mathcal{U}(t)\cos\theta(t)+\mathcal{W}'(t)$, with correction terms $\mathcal{V}'(t)$ and $\mathcal{W}'(t)$ to be determined. Then to see transitions between dressed states $|\Psi_0(t)\rangle$ and $|\Psi_{\pm}(t)\rangle$, we apply a unitary transformation $U(t)=\sum_{n=\pm,0}|\Psi_n\rangle\langle\Psi_n(t)|$ to recast Hamiltonian into a new frame. In this frame, Hamiltonian takes the form
	\begin{eqnarray}
	\hat{H}_{1}^{\prime}(t)&=& U(t)\hat{H}^{\prime}_{\rm eff}(t)U^{\dagger}(t)-iU(t)\dot{U}^{\dagger}(t)\nonumber\\
	&=&\lambda(t)(|\Psi_{+}\rangle\langle\Psi_{+}|-|\Psi_{-}\rangle\langle\Psi_{-}|)+[\eta_{+}(t)|\Psi_{+}\rangle\langle\Psi_{0}|+\eta_{-}(t)|\Psi_{-}\rangle\langle\Psi_{0}|+\mathrm{H.c.}],
	\end{eqnarray}
where the three time-dependent parameters are given by
	\begin{eqnarray}
	\lambda(t)&=&\cos\mu(t)[\mathcal{W}'(t)\cos\theta(t)-\mathcal{V}'(t)\sin\theta(t)+\mathcal{U}(t)\cos2\theta(t)]+\dot{\theta}(t)\sin\mu(t),\\
	\eta_{\pm}(t)&=& i\{\dot{\theta}(t)\cos\mu(t)-\sin\mu(t)[\mathcal{W}'(t)\cos\theta(t)-\mathcal{V}'(t)\sin\theta(t)+\mathcal{U}(t)\cos2\theta(t)]\}\nonumber\\
	&&\mp[\mathcal{V}'(t)\cos\theta(t)+\mathcal{W}'(t)\sin\theta(t)+\mathcal{U}(t)\sin2\theta(t)+\dot{\mu}(t)].
	\end{eqnarray}
Apparently, $\eta_{\pm}(t)$ characterize the coupling strengths between $|\Psi_0(t)\rangle$ and $|\Psi_\pm(t)\rangle$. Consequently, the condition $\eta_{\pm}(t)=0$ ensures that $|\Psi_0(t)\rangle$ can serve as the evolutionary path without leakages into $|\Psi_\pm(t)\rangle$. From $\eta_{\pm}(t)=0$, we readily obtain 
\begin{eqnarray}
\mathcal{V}'(t)&=& -\sin\theta(t)[\dot{\theta}(t)\cot\mu(t)+\mathcal{U}(t)]-\dot{\mu}(t)\cos\theta(t), \nonumber\\
\mathcal{W}'(t)&=& \cos\theta(t)[\dot{\theta}(t)\cot\mu(t)-\mathcal{U}(t)]-\dot{\mu}(t)\sin\theta(t),
\end{eqnarray}
which exactly result in the modified Rabi frequencies expressed by eq.~\eqref{eq14}.
In our work, we achieve the quantum state transfer from $|2,-2\rangle$ to $|2,0\rangle$ by using this STA method through evolution along $|\Psi_0(t)\rangle$ with parameters given in eq.~\eqref{e14} satisfying the conditions $\mu(0)=\mu(t)=0$, $\theta(0)=0$, and $\theta(t)=\pi/2$.
\end{widetext}

\end{document}